# Dual-polarization light emission from InAs quantum dots in a annular photonic crystal cavity

Liyong Jiang*,[1,2] Wei Jia,[1] Hong Wu,[1] Wei Zhang,[1,2] Wei Su,[1] and Xiangyin Li[1]

[1]*Nanophotonic Laboratory, Department of Physics, Nanjing University of Science and Technology, Nanjing 210094, China*
[2] *Centre for Disruptive Photonic Technologies, School of Physical and Mathematical Sciences, Nanyang Technological University, 1 Nanyang Walk, Blk 5 Level 3, Singapore 637371*
*\*Corresponding author: jly@mail.njust.edu.cn*

The annular photonic crystals have been regarded as a satisfactory candidate to realize dual-polarization photonic device. In this letter, we focus our attention on the study of annular photonic crystal cavity to verify its application in light emission. We proposed a two-dimensional photonic crystal model with annular air units and a point-line defect to construct a cavity for the enhancement of light emission of InAs quantum dots. With the help of global optimization method, we have obtained an annular photonic crystal cavity design which can show a high in-plane quality factor of about $1.3 \times 10^5$ and $2.8 \times 10^6$ for transverse electric and transverse magnetic polarizations, respectively. Based on the Electron Beam Lithography and Reactive Ion Etching techniques, such cavity pattern was transferred into the top of InAs/GaAs active layer. The photoluminescence spectra of sample demonstrated clear light emission at around 1.3 μm for both polarizations. Such dual-polarization light emitter has potential applications in optical communications, quantum information processing and other relative areas.

It is well known that the most important factor which can affect the efficiency of a quantum-dot light emitter is the control of spontaneous radiation [1]. The photonic crystal micro- or nano- cavities are an ideal platform to control the spontaneous radiation since they can support localized cavity modes with high quality factor $Q$ and low mode volume $V$, thus can generate high Purcell factor [2] $F_p = 3Q\lambda^3/4\pi^2Vn^3$ to enhance the spontaneous emission rate of the quantum dots (QDs). For instance, ultra-high $Q$ ($>1\times10^6$) has been demonstrated with a mode volume comparable to a cubicoptical wavelength [3].

In particular, in the past decade, the idea of utilizing photonic crystal nanocavities to modify light emission and enhance quantum yield from a semiconducting emitter has led to fruitful achievements in InAs QDs quantum-dot light emission with main working wavelengths in infrared range [4-19]. For example, single self-assembled semiconductor InAs QDs working at around 1.3 μm have been extensively studied since they are regarded as one of the most promising candidates for the future infrared fiber-based applications of quantum communication. In some previous works [16-19], kinds of photonic crystal cavities have been adopted to enhance the light emission from InAs QDs at around 1.3 μm. However, they were only focused on one polarization. As a complementary work, the present study will be focused on the light emission from InAs QDs at around 1.3 μm for dual polarizations based on the annular photonic crystal cavities.

The annular photonic crystals (APCs) [20] are usually constructed by merging a rod-type photonic crystal into a hole-type one, which means an annular air ring embedded in a dielectric background. As a result, compared with conventional rod-type photonic crystals, the additional dielectric background will provide extra support for transverse electric (TE) polarization modes than a pure air background. Similarly, the dielectric rods inside the air holes are sensitive to transverse magnetic (TM) polarization modes more than the pure air holes in conventional hole-type photonic crystals. Recently, APCs have been demonstrated to be an ideal candidate for realizing dual-polarization devices [21-24], including high-transmission waveguides, beam splitters, super lens, and slow-light waveguides. In this letter, we will make a further demonstration that the APCs can also be applied in quantum-dot light emission.

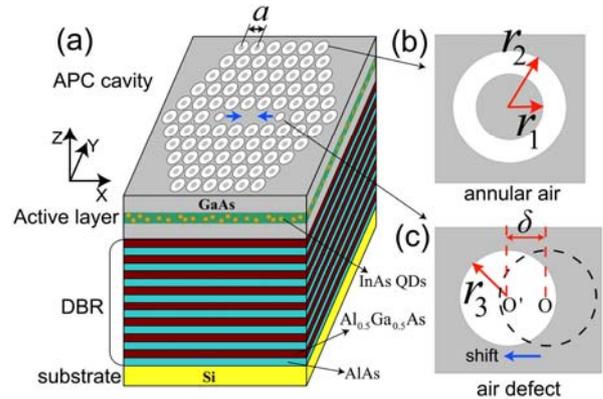

Fig. 1. (Color online) (a) Illustration of the quantum-dot light emitter model studied in this paper. The active layer is made up of GaAs with InAs QDs embedded in it. The distributed-Bragg reflector is made up of alternating $Al_{0.5}Ga_{0.5}As$ and AlAs layers with quarter-wave thickness in each layer. (b) and (c) show the annular air unit and point air defect of the annular photonic crystal (APC) cavity (L3-type) pattern on the top of active layer. $r_1$ and $r_2$ represent inner and outer radius of the annular air unit, while $r_3$ and $\delta$ represent the radius and location shift value of the air defect.

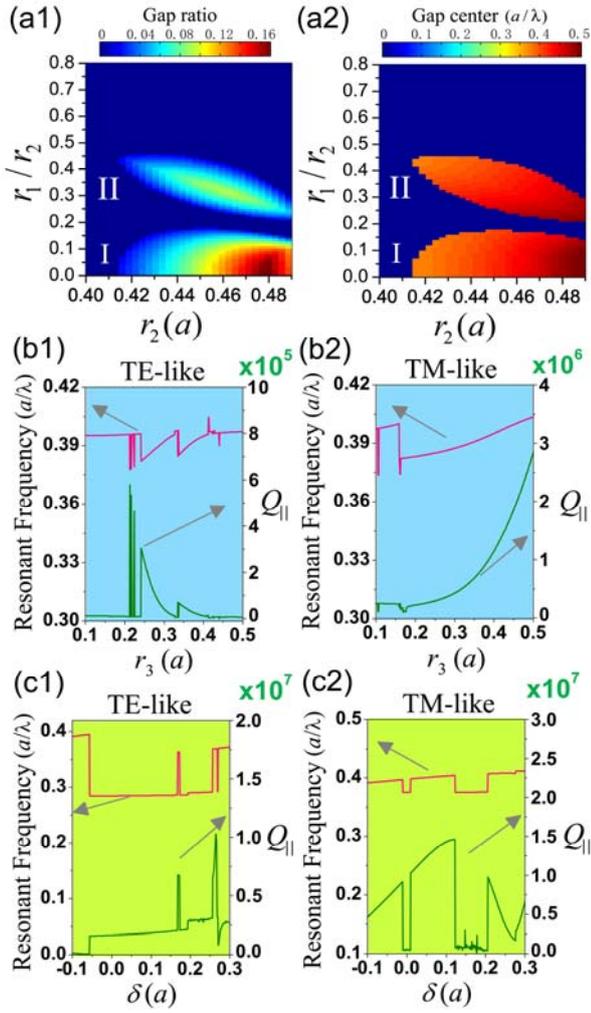

Fig. 2. (Color online) (a1) and (a2) Common band-gap scanning as the outer and inner radii of annular air unit vary. When the outer and inner radius of annular air is $0.43a$ and $0.17a$, the resonant frequency and in-plane quality factor $Q_{\parallel}$ scanning for TE-like and TM-like polarizations as the radius of point air defect (b1 and b2) and the location shift value of point air defect (c1 and c2) vary.

The entire quantum-dot light emitter is based on Silicon substrate and the structure is illustrated in Fig. 1(a). We consider self-assembled InAs QDs as light emission material and they are embedded in GaAs background. A one-dimensional Distributed-Bragg Reflector (DBR) which is constructed by alternating $Al_{0.5}Ga_{0.5}As$ and AlAs layers is used as backward reflector. In particular, to control the spontaneous radiation of InAs QDs, a two-dimensional annular photonic crystal cavity is introduced into the top InAs/GaAs layer. As shown in Fig. 1(b), the unit cell of two-dimensional photonic crystal is an annular air with inner radius $r_1$ and outer radius $r_2$. The period of triangular-lattice photonic crystal is $a$. To construct a cavity in the photonic crystal, a widely used modified L3-type defect model is considered here. Three annular-air units are removed near the center of photonic crystal to form a L3-type [25, 9-17] linear defect while another two annular-air units adjacent to the end of linear defect are replaced by air holes to form two point defects. As further shown in Fig. 1(c), the radius of air holes is represented by $r_3$ and the location of air holes is modified with a small shift value $\delta$ along the direction of line defect.

For such a model of quantum-dot light emitter, the main task in theory is focused on the design of annular photonic crystal cavity to show high quality factor $Q$ for cavity modes. In particular, we employed the plane wave expansion (PWE) method to analyze the band diagram of photonic crystal. A freely available package MEEP [26] which based on the three-dimensional finite difference time-domain (FDTD) method was used to calculate the cavity mode profile, quality factor $Q$, and mode volume $V$.

We began with numerical simulations by finding suitable values for radii $r_1$ and $r_2$ which can sensitively influence the common band gap of the annular photonic crystal for both polarizations. From Fig. 2(a1) it can be found that common band gaps will exist only when $r_2$ is larger than about $0.414a$. Under this condition, there are two sub areas which can support common band gaps when $r_1$ varies from 0 to $0.8r_2$. Area I can be observed when $r_1$ is from about 0 to $0.15r_2$ while area II is available from about $0.2r_2$ to $0.45r_2$. The central frequency of common gap is mainly focused in the range of [0.4 $a/\lambda$, 0.5 $a/\lambda$] for these two areas [seen in Fig. 2(a2)]. We can further find in area I that high gap ratios can be obtained when relative large $r_2$ and relative small $r_1$ are simultaneously satisfied, while in area II there is an elliptical core which can support high gap ratios. On the other side, considering the fabrication precision, samples with too small $r_1$ will be quite difficult to be fabricated. At the same time, the main purpose of introducing APCs into this study is to find well support for both polarizations at a specific frequency, rather than a large range of common frequencies. Since dielectric rods favor TM-like modes than TE type, it is better to use a relative large $r_1$ to well support TM-like modes. As a result, we finally chose an outer radius $0.43a$ and an inner radius $0.17a$ for further study. This annular photonic crystal shows a common band gap at central normalized frequency $0.392$ $a/\lambda$ with a gap ratio of 5.6%.

We next transferred our attentions to the optimization of point-line defect to find high $Q$ designs for both polarizations at the same frequency. In general, total $Q_{total}$ of the light emitter can be expressed as $Q_{total} = Q_{\parallel}Q_{\perp}/(Q_{\parallel}+Q_{\perp})$, where $Q_{\parallel}$ and $Q_{\perp}$ represent the parallel (in x-y plane) and vertical (out-of x-y plane) quality factor respectively. To gain high $Q_{total}$ value, the most widely used method is to suppress the in-plane loss and enhance the parallel quality factor $Q_{\parallel}$. Figs. 2(b1) and 2(b2) show the relationships between point defect radius $r_3$, resonant frequency and quality factor $Q_{\parallel}$ for different polarizations. It can be found that the main resonant frequency show similar range between around 0.38 $a/\lambda$ and 0.405 $a/\lambda$ for both polarizations as $r_3$ changes from $0.1a$ to $0.5a$. Considering the common band gap of annular photonic crystal is between 0.381 $a/\lambda$ and 0.403 $a/\lambda$., it is believed that the band gap has made a main contribution to the resonant cavity modes and it is relative easy to find common resonant frequency during this gap. However, the quality factor response is quite different for different polarizations. In general, the $Q_{\parallel}$ value is in the order of $10^5$ and for $10^6$ TE-like and TM-like polarizations respectively. Meanwhile, the $Q_∥$ value will gradually decrease for TE-like polarization while an opposite behavior can be found for TM-like polarization. So it is better to choose a middle value of $r_3$ to gain relative high $Q_∥$ value for both polarizations. We have also investigated the influences of point defect location shift value $δ$ on resonant frequency and $Q_∥$ value for different polarizations. The results are shown in Figs. 2(c1) and 2(c2). As $δ$ changes from -0.1$a$ to 0.3$a$, the resonant frequency range is about [0.286 $a/λ$, 0.395 $a/λ$] and [0.376 $a/λ$, 0.412 $a/λ$] for TE-like and TM-like polarizations, respectively. These ranges do not match well with the common band gap of annular photonic crystal, especially for TE-like polarization which shows a very wide range beyond the bottom edge of band gap. Such mismatch is believed due to the Fano resonant cavity modes in L3 line defect because such cavity modes are very sensitive to the cavity length. At the same time, compare with defect radius $r_3$, we can further find that the shift value $δ$ will more easily result in relative high $Q_∥$ values for both polarizations.

Considering above lineally scanning results are too complicated, it is difficult to find an optimum designs from these plots which can show high $Q_∥$ values for both polarizations at the same frequency during the common band gap. Instead, we treated such optimization by employing global optimization strategy. In particular, the Genetic algorithm was employed to do such global optimization task. To narrow the searching space as possible as we can, we set a target resonant frequency at 0.39 $a/λ$ and a searching space [0.2$a$, 0.4$a$] for $r_3$ and [0, 0.3$a$] for $δ$ respectively.

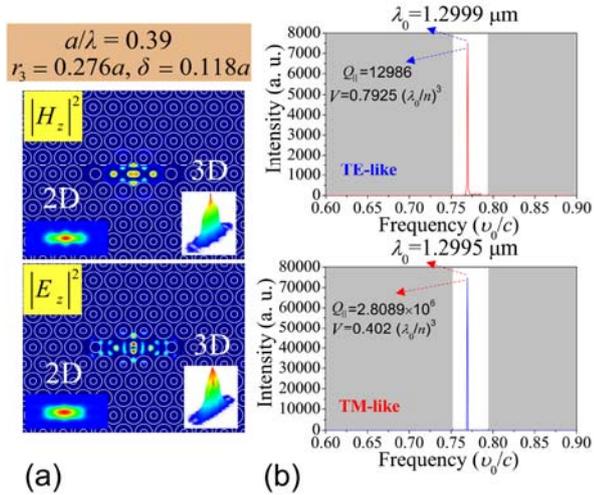

Fig. 3. (Color online) (a) The $z$-directional magnetic and electric field distributions in $x$-$y$ plane as well as the 2D and 3D far-field out-of-plane intensity profiles (shown in insets) at normalized frequency 0.39 $a/λ$ when the optimum parameters of cavity structure are $r_3$=0.276$a$ and $δ$=0.118$a$. (b) Corresponding $x$-$y$ plane light transmission spectrum of the optimum cavity structure when period constant of lattice is 507 nm. The white area indicates the range of band gap of annular photonic crystal.

The finally optimum structure for point-line defect are given as follows, $r_3$=0.276$a$ and $δ$=0.118$a$. Assuming the lattice $a$ is 507 nm, the corresponding target resonant wavelength $λ_0$ will be 1.3 μm at normalized frequency 0.39 $a/λ$. Fig. 3(a) shows the magnetic and electric field distributions in $x$-$y$ plane. We can observe that the resonant cavity mode with either TE-like or TM-like polarization is well confined in the defect and they are obviously fundamental modes from the 2D and 3D far-field out-of-plane intensity profiles (see insets). To gain more detailed insights of cavity modes, Fig. 3(b) further shows the light transmission spectrum in $x$-$y$ plane. It can be clearly found that there is a very sharp transmission peak during the band gap of annular photonic crystal. The specific peak is located at around 1.3 μm with a ultra-high $Q_∥$ value of $10^5$~$10^6$ order as well as a very small mode volume of $(λ_0/n)^3$ order for both TE-like and TM-like polarizations. The relative low $Q_∥$ value for TE-like polarization also corresponds to a relative large mode volume. This volume difference can be observed from Fig. 3(a), where the magnetic field seems to be a bit leakier than electric field.

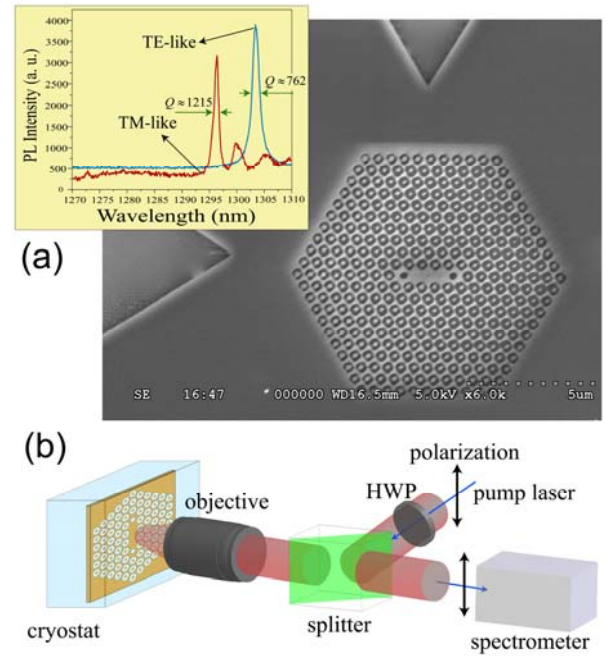

Fig. 4. (Color online) (a) The top SEM image of the fabricated quantum-dot light emitter sample. The inset shows the filtered photoluminescence spectra of the sample. (b) Simple schematic diagram of the measurement setup for photoluminescence spectra.

We started the fabrication of quantum-dot light emitter sample by firstly taking the Metal-organic Chemical Vapor Deposition technique to deposit a 25-paris quarter-wave DBR stack $(Al_{0.5}Ga_{0.5}As/AlAs)^{25}$ on the Si (111) substrate. Considering the desired working wavelength for quantum-dot light emitter is 1.3 μm, we set the thickness for $Al_{0.5}Ga_{0.5}As$ and AlAs layer as 94nm and 112 nm, respectively. Then we further deposited a 240 nm thick GaAs waveguide layer with several monolayer InAs QDs (density is about $3×10^{10}$ dots/cm$^2$) embedded at the center of it. The purpose of using high density of InAs QDs is to maximize the coupling to the fundamental cavity mode. After that, we employed the

Electron Beam Lithography (EBL) in conjunction with the Reactive Ion Etching (RIE) techniques to fabricate the optimum annular photonic crystal cavity on the top of InAs/GaAs layer. The period constant $a$ was set as 507 nm and considering the fabrication precision of EBL, the other parameters were chosen as follows: $r_1$=218 nm, $r_2$=86 nm, $r_3$=140 nm, $\delta$=60 nm. After 500 nm 950K MW PMMA resist was spin coated on the top surface of the wafer, the patterns of annular photonic crystal cavity were written with the EBL system (STS ELS-7000). Then these patterns were transferred into the top of InAs/GaAs layer using RIE etching technique. The etching was conducted in $SiCl_4$/Ar plasma and finally the remaining resist was removed with a $NH_4OH$/acetone soak.

The top scanning electron microscopy (SEM) image of the quantum-dot light emitter sample is shown in Fig. 4(a). It can be found that the precision of inner rods is not very satisfactory. Considering the precision of our EBL system is about 60 nm, the consistency of inner radius $r_1$ will be strongly influenced. Such manufacturing error is still difficult to be completely eliminated but it may be further decreased by choosing ZEP520A as an electron-beam resist and carefully optimizing the exposing time during the EBL process. We characterized the photoluminescence (PL) spectra of quantum-dot light emitter sample by using a micro-photoluminescence spectrometer with a broad-band laser excitation source. The measurement setup is shown in Fig. 4(b). The sample was mounted in a He-flow cryostat cooling box and it was pumped by a 780 nm continuum diode laser with about 2 μm spot size focused by a microscope objective (100×, NA=0.7). The emission light was collected by the same objective and detected with a cooled charge-coupled device camera in the spectrometer. A half-wave plate was used to change the direction of polarization.

The filtered PL spectra of this sample are shown in the inset of Fig. 4(a). The measurement of PL spectra was under the temperature of 4 K and a pumping power of 28 μW. It can be found that the sample can show a clear PL peak at around 1.303 μm ($Q \approx 762$, $F_p \approx 73$) and 1.295 μm ($Q \approx 1215$, $F_p \approx 230$) for TE-like (electric field parallels to $y$ direction) and TM-like (electric field parallels to $x$ direction) polarizations, respectively. The higher total $Q$ value and Purcell factor of TM-like polarization is obviously due to the higher theoretical designs. These two main peaks can demonstrate the coupling between InAs QDs spontaneous radiation and resonant cavity modes. The mismatch of measured and theoretical resonant peak locations is due to the unavoidable and non-ignorable manufacturing inaccuracies. The simulation results show that if the inner radius $r_1$ suffers a small variation of 0.01$a$, there will be about 6 nm shift of the peak location. We can also find two additional small peaks in the PL spectrum for TM-like polarization. They are believed to be caused by the coupling between InAs QDs light emission and other cavity modes in modified L3 defect. In general, the temperature-depended and time-resolved spectra are also very important for InAs QDs. Considering the manufactured accuracy of sample is not satisfactory, we did not study these photon emission properties in this letter. We will take them into account in our further study after the manufactured accuracy of APCs can be obviously improved.

To sum up, we have successfully applied the annular photonic crystal cavity into the quantum-dot light emission system to construct a dual-polarization light emitter. The simulation results have demonstrated an optimum design of annular photonic crystal cavity with high $Q_\parallel$ value for both TE-like and TM-like polarizations. The experimental results can further demonstrate clear PL spectra of InAs QDs at around 1.3 μm for both polarizations when the annular photonic crystal cavity is adopted.

This work was supported by the Natural Science Foundation of China (No: 61205042) and the Natural Science Foundation of Jiangsu Province in China (No: BK2014021828). The first author also wishes to acknowledge the financial support from the China Scholarship Council as well as the Zijin Intelligent Program of NUST (No: 2013_zj_010203_16).

**References**
1. S. Noda, M. Fujita, and T. Asano, Nat. Photonics **1**, 449 (2007).
2. E. M. Purcell, Physical Review **69**, 681 (1946).
3. T. Tanabe *et al.*, Nat. Photonics **1**, 49 (2007).
4. J. Vuckovic, and Y. Yamamoto, Appl. Phys. Lett. **82**, 2374 (2003).
5. S. Laurent *et al.*, Appl. Phys. Lett. **87**, 163107 (2005).
6. T. Yoshie *et al.*, Appl. Phys. Lett. **79**, 114 (2001).
7. D. G. Gevaux *et al.*, Appl. Phys. Lett. **88**, 131101 (2006).
8. F. S. F. Brossard *et al.*, Appl. Phys. Lett. **97**, 111101 (2010).
9. S. M. Thon *et al.*, Appl. Phys. Lett. **94**, 111115 (2009).
10. R. Bose *et al.*, Phys. Rev. Lett. **108**, 227402 (2012).
11. J. Lee *et al.*, Phys. Rev. Lett. **110**, 013602 (2013).
12. A. Faraon *et al.*, Phys. Rev. Lett. **104**, 047402 (2010).
13. H. Kim *et al.*, Opt. Express **19**, 2589 (2011).
14. L. Midolo *et al.*, Appl. Phys. Lett. **101**, 091106 (2012).
15. B. Ellis *et al.*, Nat. Photonics **5**, 297 (2011).
16. L. Balet *et al.*, Appl. Phys. Lett. **91**, 123115 (2007).
17. M. Nomura *et al.*, Phys. Rev. B **75**, 195313 (2007).
18. W. Y. Chen *et al.*, Appl. Phys. Lett. **87**, 071111 (2005).
19. W. Y. Chen *et al.*, Solid State Commun. **150**, 1798 (2010).
20. H. Kurt, and D. S. Citrin, Opt. Express **13**, 10316 (2005).
21. A. Cicek, and B. Ulug, Opt. Express **17**, 18381 (2009).
22. J. Hou *et al.*, Opt. Commun. **282**, 3172 (2009).
23. L. Y. Jiang *et al.*, J. Appl. Phys. **111**, 023508 (2012).
24. H. Wu *et al.*, Appl. Phys. Lett. **102**, 141112 (2013).
25. Y. Akahane *et al.*, Nature **425**, 944 (2003).
26. See http://ab-initio.mit.edu/mpb/ for information about MIT photonicbands.